\documentclass[smallcondensed]{svjour3}    
\smartqed  
\usepackage{graphicx}
\usepackage{amssymb}
\usepackage{amsfonts}
\usepackage{enumerate} 
\usepackage{array}
\usepackage{amsmath}
\usepackage{comment}

\usepackage{natbib}
\usepackage{url}

\newcommand\crm{\cr\noalign{\medskip}}

\newcommand\norm[1]{\left\Vert#1\right\Vert}

\newcommand\Frac[2]{{{\displaystyle\strut#1}\over{\displaystyle\strut#2}}}
\newcommand\Dron[2]{\Frac{\partial#1}{\partial#2}}

\newcommand\Der[2]{\Frac{d#1}{d#2}}                                                
\newcommand\Dt[1]{\Frac{d#1}{dt}}

\newcommand\be{\begin{equation}}
\newcommand\ee{\end{equation}}
\def\m@th{\mathsurround=0pt}
\newcommand\EQM[1]{\vcenter{\normalbaselines\m@th
    \ialign{${\displaystyle ##}$\hfil&&\ ${\displaystyle ##}$\hfil\crcr
    \mathstrut\crcr\noalign{\kern-\baselineskip}
    \noalign{\smallskip}
    #1\crcr\mathstrut\crcr\noalign{\kern-\baselineskip}}}}

\renewcommand\L{L}
\renewcommand\a{{\alpha}}

\def\vuu{{\bf u}}
\def\vrr{{\bf r}}
\def\vii{{\bf i}}
\def\vjj{{\bf j}}
\def\vpp{{\bf p}}
\def\vnn{{\bf n}}
\newcommand\vkk{{\bf k}}
\newcommand\vKK{{\bf K}}

\newcommand\vGG{{\bf G}}
\newcommand\vPP{{\bf P}}
\newcommand\dr{{\bf \dot r}}
\newcommand\ddr{{\bf \ddot r}}

\newcommand\beq{\begin{equation}}
\newcommand\eeq{\end{equation}}
\newcommand\om{\omega}
\newcommand\OM{\Omega}

\def\V{{\bf V}}
\def\cH{{\cal H}}
\def\cA{{\cal A}}

\def\vI{{\bf I}}
\def\vJ{{\bf J}}
\def\dX{\dot X}
\def\dY{\dot Y}

\def\vA{{\bf A}}
\def\vb{{\bf b}}

\journalname{Celestial Mechanics and Dynamical Astronomy}

\begin{document}

\title{Andoyer construction for Hill and Delaunay variables
}


\author
{
	Jacques Laskar   	
}


\institute{
	 Jacques Laskar \at
    ASD, IMCCE-CNRS UMR8028, Observatoire de Paris, UPMC,
    77 Av. Denfert-Rochereau, 75014-Paris, France 
	\email{laskar@imcce.fr}
 }

\date{Received: \today/ Accepted: date}

\maketitle

\begin{abstract}

Andoyer variables are well known for the study of rotational  dynamics. These 
variables were derived by Andoyer through a  procedure that can be also used 
to obtain  the Hill variables of the Kepler problem.
Andoyer construction can also  forecast the Delaunay variables 
which  canonicity  is then obtained without the use of a generating function.

\end{abstract}

\renewcommand\L{L}

\section{Introduction}
\label{intro}

Buiding on  the  work  of \citet{Bine1841a}, 
Delaunay introduced the  set of coordinates that he used for the elaboration of his Lunar 
theory  \citep{Dela1860a}. A few years later, in his thesis, 
Tisserand demonstrated how the canonicity of the Delaunay variables can  be obtained  
 through Hamilton-Jacobi theory \citep{Tiss1868a}. 
Tisserand's presentation is  now ubiquitous in celestial mechanics 
textbooks. 
Here we present a derivation of the Delaunay variables that does not require Hamilton-Jacobi theory. 
We use the intermediate derivation of the Hill variables \citep{Hill1913a} following 
Andoyer \citep{Ando1915a,Ando1923a}. Transformations from Hill variables to Delaunay variables 
exist in  the literature \citep{Ando1913a,Depr1981b,Flor1995a}  but they rely on a 
generating function  which we will avoid here by using a 
 direct computation based on the invariance of the canonical differential 2-form.
 
Henri Andoyer (1862-1929)  is well-known  for the derivation of the Andoyer variables 
that are very well adapted to rotational dynamics \citep{Ando1915a,Ando1923a}. In fact,  the derivation 
of these variables is obtained through a very general procedure that can also be applied to the 
Keplerian two-body problem, and which then leads to the Hill variables \citep{Ando1915a,Ando1923a}.

\section{Andoyer canonical criterion}
We recall here the criterion for canonicity of \citet{Ando1923a}. Let us consider a $2n$ dimensional  
 phase  with coordinates the canonical variables $(p_j,q_j)$, where $q_j$ are coordinate-type variables and 
$p_j$ the momenta. We then make the change of variables $(p_j,q_j) \longrightarrow (y_k, z_k)$
which we assume to be a good differentiable change of variables on the domain of interest, 
but without any assumption on its canonicity.  We have
 \be
 \EQM{
 \sum_j p_j\, dq_j &= \sum_{j} p_j\left(\sum_{k} \Dron{q_j}{y_k}dy_k \right)+ \sum_{j}p_j\left( \sum_{l} \Dron{q_j}{z_l}dz_l \right)\crm
 & = \sum_{k} \left(\sum_{j} p_j\Dron{q_j}{y_k}\right) dy_k + \sum_{l}\left(\sum_{j} p_j \Dron{q_j}{z_l}\right)dz_l \crm \ .
 }
 \label{dform}
 \ee
 
 A change of variable is canonical if it preserves the 2-form $\sigma_2 = \sum_j dp_j \wedge dq_j $ \citep[e.g.][]{Arnold1989a}. 
This will be in particular the case if the change of variable preserves the 1-form $\sigma_1=\sum_j p_j  dq_j $,  as $d(\sigma_1)=\sigma_2$.
Thus, the  transformations  preserving the 1-form $\sum_j p_j\, dq_j $ form a subgroup of the canonical 
transformations. They  are called Mathieu canonical transformations \citep{Math1874a,Whit1904a}. 
 For any variable $\a$, let us denote 
 \be
J_\a = \sum_{j} p_j \Dron{q_j}{\a} \ .
\ee
 Equation (\ref{dform}) thus becomes 
 \be
 \sum_j p_j\, dq_j  = \sum_{k} J_{y_k} dy_k  +  \sum_{l} J_{z_l} dz_l \ .
 \ee
 
 Andoyer assumes that for all $k=1,\dots,n $,  $J_{y_k} = 0$, 
 and that for all $ l=1,\dots,n$,  $J_{z_l} =  u_l(y)$, 
where $(y_k)_{k=1,\dots,n} \longrightarrow (u_k)_{k=1,\dots,n} $ is a diffeomorphism. We have then 
 \be
 \sum_j p_j\, dq_j = \sum_j u_j\, dz_j  \ .
 \ee
 The change of variables $(p_j,q_j)_{j=1,\dots,n} \longrightarrow (u_j, z_j)_{j=1,\dots,n}$  preserves the 
  1-form $\sum_j p_j\, dq_j $ and therefore preserves also the canonical 2-form 
$
\sum_j dp_j \wedge dq_j \ .
$
It is thus   canonical.
To search for such a canonical change of variables, 
one thus needs to compute $J_\a$ for any variable $\a$ in the set $\{y_k,z_k\}_{k=1,\dots ,n}$.
Andoyer remarks then that  $J_\a$ is the scalar product
\be
J_\a = \vpp \cdot  \V _\a  \ ,
\label{JJ}
\ee
where  $\vpp =(p_j)$ is the momentum vector, and the virtual velocity $\V_\a =\left(\Dron{q_j}{\a}\right)$ is   
obtained when  only varying the variable $\a$ in the position vector $(q_j)$. 
When the Andoyer criterion is verified, the quantities $J_\alpha$ will  be the new conjugate 
momentum associated to the coordinate-type variables $\alpha$. 

\section{Hill variables}
\label{sechill}

For a celestial body $P$, we consider the Kepler problem in a fixed reference frame $(\vii,\vjj,\vkk)$, 
whose origine $O$ coincides with the attractive center, 
with radius vector $\vrr=r \vuu$, velocity $\dr$,
gravitational coupling parameter $\mu$, and
Hamiltonian per unit of mass
\be
\cH = \Frac{1}{2} \dr^2 -\Frac{\mu}{r} \ .
\label{kep}
\ee

\begin{figure}[h]
\includegraphics[width=7cm]{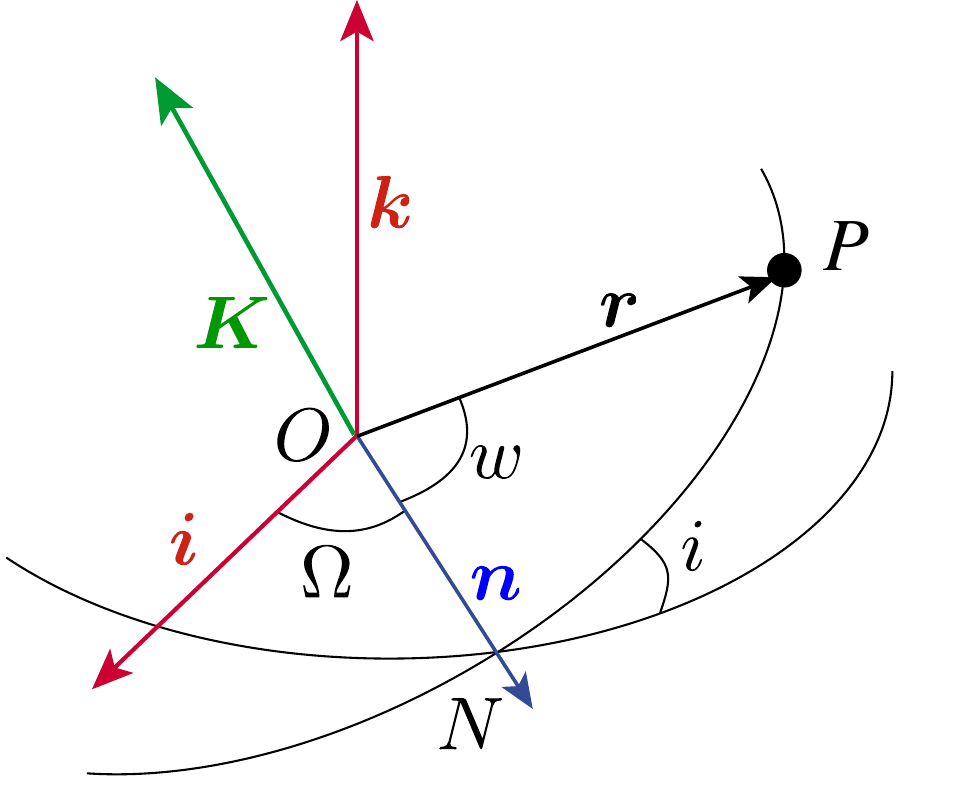} 
\caption{Reference frame and notations. The vectors $\vii, \vkk, \vKK, \vnn$  are unit  vectors. $N$ is the ascending node. 
}
\label{fig1} 
\end{figure}

The orbital plane $(\vrr,\dr)$ is orthogonal to the angular momentum per unit mass $\vGG=G \vKK$ ($G=\norm{\vGG}$), and is defined 
by the longitude of the ascending node $\Omega$ and inclination $i$ (Fig.1). The position of the celestial body $P$ 
is  defined when $r$ and  the argument of   latitude $w = \omega + v$ ($\omega$ is the argument of perihelion, and $v$ the true  
anomaly) are given. We have thus characterized the position of $P$ with the four variables $(r,\Omega,w,i)$. With Hamiltonian (\ref{kep}), the momentum 
vector is simply the velocity $\dr$. Following Andoyer, 
we  extend the  transformation $ (\dr,\vrr) \longrightarrow (r,\Omega,w,i)$ into a genuine canonical change of variables.
For this, we evaluate all  $\V_\a$ and $J_\a =\dr \cdot \V_\a$ quantities (\ref{JJ}).
We remind  that when rotating a vector  $\vA$ around a fixed unit vector  $\vb$  by an angle  $\theta$, we have 
\be
\Der{\vA}{\theta}= \vb \wedge \vA \ .
\ee

The virtual velocity $\V_i$ is obtained from a rotation   of angle $i$ around  $\vnn$, the unit vector in the direction of ${\bf ON} $ (Fig.1). 
We have thus 
\be
 \V_i = \vnn \wedge \vrr \ , \qquad J_i = \dr \cdot \V_i = 0 \ .
 \label{hilleqa}
 \ee
In a similar way, we have 
\be
\EQM{
 \V_w= \vKK \wedge \vrr \ ,\qquad  &J_w &= \dr \cdot (\vKK \wedge \vrr) = \vKK \cdot (\vrr \wedge \dr) = G \ , \crm
 \V_\Omega= \vkk \wedge \vrr \ ,   &J_\Omega &= \dr \cdot (\vkk \wedge \vrr) = \vkk \cdot (\vrr \wedge \dr) = G\cos i \ .
 \label{hilleqb} }
\ee
and as   $\vrr = r \vuu$ 
\be
 \V_r = \vuu \ , \qquad  J_r = \dr \cdot \vuu =(\dot r\vuu + r \dot\vuu)\cdot \vuu = \dot r \ .
 \ee
Moreover, as $\vrr$ depends only on  $r,w,\Omega, i$ and not on  $\dot r$ or $G$, 
we have  $\V_{\dot r}=\V_G=0$, and then   $J_{\dot r}=J_G=0$. We are thus in the framework of the application of 
Andoyer criterion and we can conclude that the change of variables 
\be
(\dr;\vrr) \longrightarrow (\dot r, G, G\cos i ; r, w, \Omega)
\ee
is canonical\footnote{
An alternate approach following  S. Breiter (2016, pers. comm.) and inspired by \citep{Depr1993a} could be purely analytical, 
with a direct computation of the canonical 1-form, gathering  the derivations of equations (\ref{hilleqa},\ref{hilleqb}). 
Indeed, we have (Fig.1)
$$
d\vrr = \vuu dr  + (\vkk \wedge \vrr) d\Omega + (\vnn \wedge \vrr) di +(\vKK \wedge \vrr) dw    \ ,
$$
thus 
$$
 \dr\, d\vrr = \dot r  \, dr + G \, dw + G\cos i \, d\Omega \ .
$$
}. 
In these new variables, known as the Hill variables  \citep{Hill1913a}, 
 the Hamiltonian becomes\footnote{
 As $\vuu$ is a unit vector, $\vuu \perp  \dot\vuu$, and thus $\dr^2 = \dot r^2 + r^2 \norm{\dot\vuu}^2$, and 
 $G = r^2 \norm{\dot\vuu} $.
 } 
\beq
\cH = \Frac{1}{2}( \dot r^2 + \Frac{G^2}{r^2} ) - \Frac{\mu}{r}.
\label{Hill}
\eeq
The Hamiltonian (\ref{Hill}) has  only one degree of freedom $(\dot r , r)$ (and one parameter $G$). It is thus obviously integrable 
but not in a simple way. To fully achieve the reduction of the problem to a trivially integrable 
Hamiltonian, we will use the Delaunay variables. 

\section{Delaunay variables}
We assume  that we have already obtained the classical orbital elements $(a,e,i,M,\omega,\Omega)$ for an elliptical orbit, 
with the above definitions for $i,\omega, \Omega$ (Fig.1).   The derivation  of  $(a,e,M)$ is given in  the Appendix.  
The mean anomaly $M$ is the angle that is proportional to the area swept by $\vrr$ from perihelion. In the two-body problem, 
we will thus  have $dM/dt = n$, where $n=2\pi/T$ is the mean motion (see the Appendix)
with $n^2a^3=\mu$.
The variable $L$ was introduced by \citet{Dela1860a}. For an elliptical orbit, we have  $\cH=-\mu/2 a$. $L$  is 
therefore defined as  the only\footnote{Up to an additive constant.}  function $L(a)$ 
such that 
\be
\Dt{M} = n =  \Der{\cH}{L} = \mu^{1/2}a^{-3/2} \ .
\ee
which implies 
\be
\Der{L}{a}= \Frac{\sqrt{\mu}}{2\sqrt{a}}, 
\ee
and thus $ L=\sqrt{\mu a}$, up to an additive  constant.  
The mean anomaly $M$ defines the  body position from perihelion. 
We will  add $\omega$ and $\Omega$ as angular variables to define the position of the perihelion.
Following the previous section, 
we can use Andoyer's ideas to complete the change of variables into a canonical change of variables. 
For $M$ fixed, varying $\omega$ is equivalent of varying $w$, 
and thus, as in section \ref{sechill}, we have $J_\omega= G$. Therefore, 
$G$ is naturally conjugate to $\omega$, and as previously for Hill variables, $J_\Omega= H$.
We have thus obtained 
the Delaunay variables in a natural way, following Andoyer ideas. We have 
\be
\EQM{
\L &= \sqrt{\mu a} \ ,\cr
G  &=\L\sqrt{1-e^2} \ ,\cr
H  &=G\cos i \ ,
}
\qquad
\EQM{
M   \ , \cr
\om  \ ,  \cr
\OM   \ .
}
\label{delaunay}
\ee

To prove that  the change of variable from Hill variables to Delaunay variables is symplectic, 
we consider the  1-form

\be
\sigma = \dot r \,dr + G\,dw + H \,d\Omega   - (L\,dM + G\,d\omega + H \,d\Omega )  \ ,
\ee 
which reduces to (with $w=v+\omega$) 
\be
\sigma =\dot r\, dr +L\sqrt{1-e^2}\, dv -  L\,dM  \ . 
\ee

In order to evaluate this expression, we use 
 the differential relations  of the Appendix (Eq.\ref{diffell}) which allow to express  $\sigma$ in 
 term of $E,e,a,dE, de, da$ as 
 \be
  {\sigma} =  2 e \sqrt{\mu a} \cos E \, dE + 2 \sqrt{\mu a} \sin E \, de +\sqrt{\Frac{\mu }{a}} \, e \sin E \, da    \ .
  \label{sig}
 \ee

The 1-form $\sigma$ is not  null. So the change of variable from  Hill variables (and thus as well from cartesian coordinates) 
 to  Delaunay variables is  not a Mathieu transformation. Nevertheless,  a simple derivate  of (Eq.\ref{sig}) gives 
 \be
 d \sigma = 0 \ . 
 \ee
Thus 
\be
d \dot r \wedge dr + dG \wedge dw + dH \wedge d\Omega   = dL\wedge dM + dG \wedge d\omega + dH \wedge d\Omega  \ , 
\ee
and the change of variable to Delaunay variables is canonical. In these variables $(L,G,H,M,\omega,\Omega)$, the Hamiltonian
(\ref{kep}, \ref{Hill}) becomes trivially integrable as
\be
\cH = - \Frac{\mu^2}{2L^2} \ .
\ee

\section{Discussion}

The Andoyer construction exposed here allowed to introduce in a natural way the Delaunay variables. 
The variables $G$ and $H$ appear as the  conjugate variables to  $\omega$ and $\Omega$ through Andoyer derivation. 
This approach can be compared to the derivation of \citep{Bili1943a, Brou1978a} which share features with Andoyer's approach. 
The main difference, apart from using the Hill variables as an intermediary step here, 
 is the absence of the need 
for generating function by relying on the conservation of the canonical 2-form\footnote{
\citet{Brou1978a} shows that $\sigma= 2d(r\dot r)$ in (\ref{sig}), and thus $2r\dot r$ can be considered as a generating function 
for the transformation to Delaunay variables. This  adds some information, but it is not clear how 
this generating function can be obtained in a natural way from (\ref{sig}).}.
Finally, it should be said that other authors have searched for derivations of the Delaunay variables that avoid 
Hamilton-Jacobi theory or generation functions. \citet{Brou1961a} compute directly all the Lagrange brackets of the 
Delaunay variables while more geometrical approaches are provided by \citep{Chan2003a,Fej2013a}.

\small
\begin{acknowledgements}
The author  thanks A. Albouy and L. Flor\'ia for historical insights, and S. Breiter for pointing out 
some inconsistency in the first version of this paper.  
\end{acknowledgements}

\section*{Appendix : Differential relations in elliptical elements}
In order to be self-consistent, here are briefly reminded some derivation of the classical elliptical elements
that are used in the present work.
The Hamiltonian of the two-body problem is (\ref{kep})
\be
\cH = \Frac{1}{2} \dr^2 -\Frac{\mu}{r} \ ,
\ee
with the associated Newton equation 
\be 
\ddr = -\nabla_{\vrr} \cH = -\mu\Frac{\vrr}{r^3} \ .
\ee 
By direct computation, we have $d\cH/dt=0$, $d\vGG/dt = 0 $ where $\vGG=\vrr\wedge \dr$ is the angular momentum, 
  ans  also $d\vPP/dt = 0$, where $\vPP$ is  the Laplace-Runge vector 
\be
\vPP = \Frac{\dr\wedge \vGG}{\mu} -\vuu \ . 
\ee
 where  $\vuu$ is the unit vector  $\vrr/r $ of angle $v$ with $\vPP$. We have thus 
 \be
 \vPP\cdot\vuu = e \cos v = \Frac{\vrr\cdot(\dr\wedge \vGG)}{r\mu} - 1 = \Frac{G^2}{\mu r }-1  \ ,
 \ee 
 where $e$ (the eccentricity) is the norm of $\vPP$. With $e<1$, we thus obtain that the orbit is an ellipse 
 of semi-major axis $a$, and eccentricity $e$, with $G=\sqrt{\mu a(1-e^2)}$ and polar equation 
 \be
 r  = \Frac{a^2(1-e^2)}{1 + e \cos v}  \ .
 \label{ellipse}
 \ee 
 As $\cH $ is constant, we can compute its value for $v=0$. We have then $\dot r=0$, $r=a(1-e)$,  and thus $\cH = -\mu/2a$. 
 The area  $\cA$ of the ellipse is 
 \be
  \cA = \pi a^2\sqrt{1-e^2} = \int_0^{2\pi} \Frac{1}{2} r^2 dv = \int_0^T \Frac{1}{2} G dt = \Frac{GT}{2} \ ,
 \ee
 where $T$ is the period of the motion. The mean motion $n=2\pi/T$ then verifies the third Kepler's law 
$n^2a^3=\mu$.
We need also to introduce the eccentric anomaly $E$ defined as the angle $FOP'$  where $P'$ is  the point on the circle 
of radius $a$ obtained by affininity  of ration $1/\sqrt{1-e^2}$ from the ellipse (Fig.\ref{kep}). 

\begin{figure}[h]
\includegraphics[width=7cm]{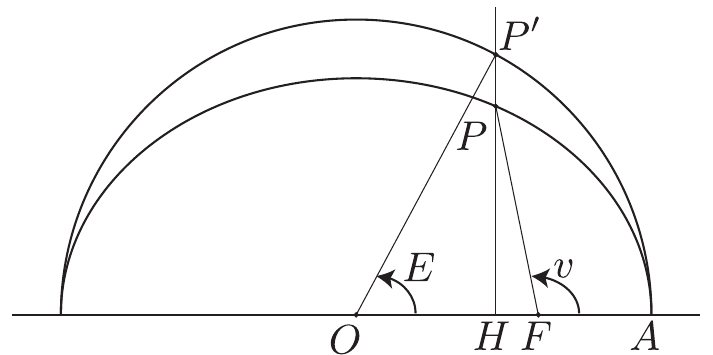} 
\caption{The circle is obtained from the ellipse by affinity of ration $1/\sqrt{1-e^2}$ from the ellipse. $v$ is the true anomaly, and $E$ 
the eccentric anomaly.
}
\label{kep} 
\end{figure}
If we denote $M$ (the mean anomaly) the angle proportional to the area $\cA(AFP)$, we have $\cA(AFP') = \cA(AFP)/\sqrt{1-e^2}=a^2M/2$.
the relation among areas $\cA(AOP') = \cA(AFP') + \cA(FOP') $ then gives the Kepler equation 
\be 
M=E-e\sin E \ .
\label{kepeq}
\ee 

In the reference frame $F, \vI,\vJ $, with $\vI$ the unit vector along $\vec{FA}$, the position and velocity $(\vrr,\dr)$ have coordinates 
$(X,Y), (\dX,\dY)$ with 
\be
\left\{ 
\EQM{
X  &= a(\cos E-e) \ ,\crm
Y &=a\sqrt{1-e^2}\sin E \ ,
}\right. 
\qquad
\left\{ 
\EQM{
\dX  &= -\Frac{\sqrt{\mu a}}{r}\sin E  \ ,\crm
\dY &=\Frac{\sqrt{\mu a}}{r}\sqrt{1-e^2}\cos E  \ .
}\right. 
\ee
with 
\be
r = a(1-e\cos E)  \ .
 \label{eqr}
 \ee
 As $\vrr\cdot\dr= r\dot r$, we have also 
 \be
 \dot r =  \Frac{\sqrt{\mu a}}{r} e \sin E \ .
 \label{eqrp}
 \ee

All relations (\ref{ellipse}, \ref{kepeq}, \ref{eqr}, \ref{eqrp}) are geometrical relations in the phase space of $(\vrr, \dr)$.
All quantities $v,M,r,\dot r$ can be expressed in term of $E, e, a$.  
By simple differentiation, we obtain the differential relations 
\be
\EQM{
dM &= \Frac{r}{a}\, dE  -\sin E \,de   \ ,\crm
dv &= \Frac{a}{r}\sqrt{1-e^2}\, dE + \Frac{a}{r\sqrt{1-e^2}} \sin E\, de  \ , \crm
dr &= ae\sin E\, dE -a \cos E\, de + \Frac{r}{a}\, da  \ ,\crm 
\label{diffell}
} 
\ee
with $r/a=1-e\cos E$.
\bibliographystyle{spbasic}      
\bibliography{andoyer}   

\begin{thebibliography}{18}
\providecommand{\natexlab}[1]{#1}
\providecommand{\url}[1]{{#1}}
\providecommand{\urlprefix}{URL }
\expandafter\ifx\csname urlstyle\endcsname\relax
  \providecommand{\doi}[1]{DOI~\discretionary{}{}{}#1}\else
  \providecommand{\doi}{DOI~\discretionary{}{}{}\begingroup
  \urlstyle{rm}\Url}\fi
\providecommand{\eprint}[2][]{\url{#2}}

\bibitem[{Andoyer(1913)}]{Ando1913a}
Andoyer H (1913) Sur l'anomalie excentrique et l'anomalie vraie comme
  {\'e}l{\'e}ments canoniques du mouvement eliptique, d'apr{\`e}s {MM. T.
  Levi-Civita et G.-W. Hill}. Bulletin Astronomique, S{\'e}rie I 30:425--429

\bibitem[{Andoyer(1915)}]{Ando1915a}
Andoyer H (1915) Sur les probl{\`e}mes fondamenteaux de la m{\'e}canique
  c{\'e}leste. Bulletin Astronomique, Serie I 32:5--18

\bibitem[{Andoyer(1923)}]{Ando1923a}
Andoyer H (1923) Cours de m\' ecanique c\' eleste, Vol. I. Gauthier-Villars,
  Paris

\bibitem[{Arnol'd(1989)}]{Arnold1989a}
Arnol'd VI (1989) Mathematical Methods of Classical Mechanics. Springer

\bibitem[{Bilimovitch(1943)}]{Bili1943a}
Bilimovitch A (1943) {\"U}ber die {Anwendungen} der {Pfaffschen} {Methode} in
  der {St{\"o}rungsthorie}. Astronomische Nachrichten 273:161

\bibitem[{Binet(1841)}]{Bine1841a}
Binet M (1841) M\' emoire sur la variation des constantes arbitraires dans les
  formules g\' en\' erales de la dynamique, et dans un syst\` eme d'\'
  equations analogues plus \' etendues. Journal de l'Ecole Polytechnique 28,
  T.XVII:1--94

\bibitem[{Broucke(1978)}]{Brou1978a}
Broucke R (1978) On {Pfaff}'s equations of motion in dynamics - {Applications}
  to satellite theory. Celestial Mechanics 18:207--222

\bibitem[{Brouwer and Clemence(1961)}]{Brou1961a}
Brouwer D, Clemence G (1961) Methods of Celestial Mechanics. Academic Press,
  New York

\bibitem[{Chang and Marsden(2003)}]{Chan2003a}
Chang D, Marsden J (2003) Geometric derivation of the delaunay variables and
  geometric phases. Celestial Mechanics and Dynamical Astronomy 86:185--208

\bibitem[{Delaunay(1860)}]{Dela1860a}
Delaunay C (1860) Th\' eorie du mouvement de la {Lune}, {Vol. I.} M\' emoires
  de l'Acad\' emie des Sciences de l'Institut Imp{\'e}rial de France
  XXVIII:1--883

\bibitem[{Deprit(1981)}]{Depr1981b}
Deprit A (1981) The elimination of the parallax in satellite theory. Celestial
  Mechanics 24:111--153

\bibitem[{Deprit and Elipe(1993)}]{Depr1993a}
Deprit A, Elipe A (1993) Complete reduction of the {Euler}-{Poinsot} problem.
  Journal of the Astronautical Sciences 41:603--628

\bibitem[{F{\'e}joz(2013)}]{Fej2013a}
F{\'e}joz J (2013) On action-angle coordinates and the {Poincar{\'e}}
  coordinates. Regular \& Chaotic Dynamics 18(6):703--718

\bibitem[{Floria(1995)}]{Flor1995a}
Floria L (1995) A simple derivation of the hyperbolic {Delaunay} variables. The
  Astronomical Journal 110:940

\bibitem[{Hill(1913)}]{Hill1913a}
Hill GW (1913) Motion of a system of material points under the action of
  gravitation. The Astronomical Journal 27:171--182

\bibitem[{Mathieu(1874)}]{Math1874a}
Mathieu E (1874) M{\'e}moire sur les {\'e}quations diff{\'e}rentielles
  canoniques de la m{\'e}canique. Journal de Math{\'e}matiques Pures et
  Appliqu{\'e}es (Journal de Liouville) XIX:265--306

\bibitem[{Tisserand(1868)}]{Tiss1868a}
Tisserand F (1868) Exposition, d'apr{\`e}s les principes de Jacobi, de la
  m{\'e}thode suivie par {M. Delaunay} dans sa th{\'e}orie du Mouvement de
  translation de la {Lune}, Th{\`e}se de doctorat. Gauthier-Villars (Paris)

\bibitem[{Whittaker(1904)}]{Whit1904a}
Whittaker E (1904) A treatise on the analytical dynamics of particles and rigid
  bodies. Cambridge Univ. Press, London

\end{thebibliography}

\end{document}